\documentstyle[prb,preprint,aps]{revtex}

\begin{document}
\title{Electron-phonon coupling close to a metal-insulator transition in one
dimension}
\author{M. Fabrizio}
\address{Istituto Nazionale di Fisica della Materia, I.N.F.M.}
\address{and International School for Advanced Studies, Via Beirut 2-4,
34013 Trieste, Italy}
\author{C. Castellani and C. Di Castro}
\address{Universit\`a di Roma ``La Sapienza'', P.le Aldo Moro 2,
00185 Roma, Italy}
\maketitle
\begin{abstract}
We consider a one-dimensional system
of electrons interacting via a short-range repulsion and coupled to
phonons close to the metal-insulator transition at half filling.
We argue that the metal-insulator transition can be described as a standard
one dimensional incommensurate
to commensurate transition, even if the electronic system is coupled to the
lattice distortion. By making use of known results for this
transition, we prove that low-momentum phonons do not play any relevant
role close to half-filling, unless their coupling to the electrons
is large in comparison with the other energy scales present in the
problem. In other words the effective strength of the low-momentum transferred
electron-phonon coupling does not increase close to the metal-insulator
transition, even though the effective velocity of the mobile carriers
is strongly diminished.

\end{abstract}
\section{Introduction}
\label{sec:Introduction}
The behavior of one-dimensional (1D) electronic systems  coupled to
phonons is a well studied problem especially in the context of the
physics of quasi-1D compounds (see e.g. Ref.~\onlinecite{review}).
Apart from their immediate applications to these materials, the 1D models
might be helpful to understand how phonons may affect higher dimensional
strongly-correlated electronic systems like, for instance, systems
close to a metal-insulator transition (MIT).
A well known example would be a model close to an incommensurate to
commensurate transition (e.g. the Hubbard model close to half
filling). A main question in this case is how the vanishing velocity
of the mobile carriers close to the MIT influences the electron-phonon
coupling. If the renormalization of the
electron-phonon coupling induced by the proximity to the MIT were
negligible, then one would conclude that the closer is the MIT,
i.e. the smaller is the quasi-particle velocity, the stronger is the
effect of the
electron-phonon coupling. This in turn might lead to anomalous but appealing
scenarios where for instance superconductivity is favored by
approaching the MIT.
The other possibility is that the MIT induces a strong
suppression of the phonon-mediated electron-electron interaction
so that no additional features
are introduced by the electron-phonon coupling if its strength is not
particularly large. At least in one dimension it is possible to give a
definite answer to the above question, which is what we are going to do
in this paper\cite{Claudio}.

Most of the theoretical works on electron-phonon coupling in 1D
deal with large momentum phonons ($q\sim 2k_F$), since, due to the nesting
properties of the 1D Fermi surface, these
phonons get strongly coupled to the $2k_F$ charge density wave
fluctuations\cite{review}.
Recently Martin and Loss\cite{Martin} have considered the different case
of low-momentum acoustic phonons coupled to electrons described
by a 1D Hubbard Hamiltonian. By adding the electron-phonon coupling
to the Luttinger-liquid model\cite{Luttinger}
which describes at low energy the Hubbard model,
they argue that, close to half filling, the system is always unstable to
a whatever small phonon-induced attractive interaction. Near this
instability, spin density wave (SDW) and charge density wave (CDW) fluctuations
are strongly depressed
and the system is pushed through a metallic phase (which was
originally found in Ref.~\onlinecite{Lucchini})
towards a superconducting phase and finally into phase separation
(the so-called Wentzel-Bardeen singularity). What they find
is therefore a point in favor of the scenario where the phonon-mediated
interaction is not strongly renormalized close to the MIT and this in turn
implies that the electron-phonon coupling constant is always
strong if compared to the effective bandwidth of the mobile carriers.
This situation could possibly occur even in higher dimensions where the
bandwidth reduction due to correlations also appears.

This result is however puzzling firstly because a similar behavior
is not shown by any physical system (e.g. doped polyacetylene) and
secondly because the tendency towards superconductivity which they find close
to half filling is difficult to reconcile with the experimental and
theoretical evidence of a Peierls phase at half filling (although one may
invoke for this purpose the region of phase separation).
At first sight, a possible explanation of this questionable result could
be the
neglect of the $2k_F$-phonons, which are known to be essential to
describe the behavior of these systems. However one easily realizes that
this is not the solution to the puzzle, since the $2k_F$-phonons would not
change the result of Ref.~\onlinecite{Martin}. In fact, let us
consider for instance the extreme situation of strong electron-phonon
coupling. In this case it is known that the lattice
distortion (i.e. the $2k_F$-phonon field) moves with the $2k_F$-CDW
fluctuations\cite{LR&A,Pietronero,Fukuyama}.
The system is then described by the
same type of Luttinger-liquid Hamiltonian\cite{Luttinger}
as in the absence of electron-phonon coupling, although with
renormalized parameters. On the other hand, charge modes described
by a Luttinger-liquid model is also the starting point of
Martin and Loss's calculation, which therefore can be redone even
in this limiting situation of
electrons strongly coupled with $2k_F$-phonons.
As a consequence their main conclusions remain valid in so far as
the Hubbard model plus the $2k_F$-phonons has a metal-insulator transition at
half filling. That their results do not depend on the neglect of
large momentum phonons is even more transparent if the $2k_F$-phonon
frequency is comparable to the electron bandwidth.
Then the electron-electron interaction
induced by the $2k_F$-phonons can be taken as unretarded and simply
renormalizes the on-site repulsion (to higher value since it is
effectively repulsive at large transferred momentum)\cite{notaa}.

In this paper we re-analyze this problem in a more systematic way.
We find, contrary to Martin and Loss, that the re-normalization of
the low-momentum transferred phonon-mediated interaction is very important
close to the MIT. The effect is to diminish the value of the coupling
constant to such an extent that the ratio of the {\sl bare}
electron-phonon coupling to the {\sl bare} bandwidth still determines the
effective strength of the electron-phonon interaction, irrespective of the
filling. When applied to the Hubbard model, our result implies that,
unless the coupling constant between the electrons and the low-momentum
phonons is sufficiently strong in comparison with the {\sl bare} hopping,
nothing changes in the phase diagram and, as the system approaches
half filling, a transition occurs from a metallic phase with
dominant density-wave fluctuations to an insulating one.

\section{The Wentzel-Bardeen singularity}
Let us consider a 1D system of interacting electrons
close to the metal-insulator transition, e.g. the Hubbard model close to
half filling, in the presence of electron-phonon interaction.

We start summarizing some known results on the low energy behavior of the
Hubbard model. The main important feature is that there are
no coherent single-particle excitations (even in the metallic phase away from
half filling), but only collective charge and spin
sound modes, which are dynamically independent. These collective fluctuations
can be imagined as sound waves in an elastic string. In a different language,
they both can be
described in terms of the Luttinger liquid phenomenology.

A Luttinger liquid has the following Hamiltonian\cite{Luttinger,Haldane-Heis}:
\begin{equation}
\hat{H}= \frac{v_\lambda}{2}\int dx\, \left\{ K_\lambda \Pi_\lambda^2(x)
+\frac{1}{K_\lambda}\left(\nabla \phi_\lambda(x)\right)^2\right\},
\label{lutt-liquid}
\end{equation}
where $\Pi_\lambda$ is the momentum conjugate to the bosonic
field $\phi_\lambda$,
and the latter are related to the
charge ($\lambda =\rho$) or spin ($\lambda =\sigma$) densities by the relations
\begin{eqnarray*}
\rho_{\uparrow}(x)+
\rho_{\downarrow}(x) &=&
\sqrt{\frac{2}{\pi}}\nabla \phi_\rho(x)
+ \frac{2}{\pi\alpha}\sin\left(2k_F x + \sqrt{2\pi}\phi_\rho(x)\right)
\cos\left( \sqrt{2\pi}\phi_\sigma(x)\right), \\
\rho_{\uparrow}(x)-
\rho_{\downarrow}(x) &=&
\sqrt{\frac{2}{\pi}}\nabla \phi_\sigma(x)
+ \frac{2}{\pi\alpha}\cos\left(2k_F x + \sqrt{2\pi}\phi_\rho(x)\right)
\sin\left( \sqrt{2\pi}\phi_\sigma(x)\right),
\end{eqnarray*}
being $\rho_{\uparrow}(x)$ and $\rho_{\downarrow}(x)$ the densities of spin-up
and spin-down electrons respectively, and
$\alpha$ an ultraviolet cut-off of the order of the lattice spacing.
$v_\lambda$ are the velocities of the charge and spin sound waves,
while the positive parameters $K_\lambda$ are related to the exponents
which characterize the asymptotic power law decay of all the correlation
functions (for non-interacting fermions $K_\lambda=1$, and for
spin isotropic interactions $K_\sigma=1$).

For the case of the Hubbard model, it is known that approaching
the metal-insulator transition, i.e. for filling $\nu\to 1/2$,
$v_\rho \to 0$ while $K_\rho \to 1/2$\cite{Schulz-Hubbard}.
As a consequence the charge compressibility $\chi\propto K_\rho/v_\rho $
diverges. On the contrary, the
spin sector is almost unaffected by the proximity to
half-filling. Exactly at $\nu=1/2$, the charge sector acquires a gap
and the correlation functions behaves as if
the effective $K_\rho=0$. In other words $K_\rho$ jumps discontinuously
from 1/2 to 0.

Now let us add to the on-site repulsion $U$ the electron-phonon
coupling $g(x-y)$ via the term
\begin{equation}
 \int dx\, g(x-y) \rho (x) u(y),
\label{el-ph}
\end{equation}
where $u(y)$ is the phonon displacement field.
By integrating out the phonons,
(\ref{el-ph}) generates  in the action $S$ a retarded electron-electron
interaction via phonon exchange, given by:
\begin{equation}
\delta S = - \frac{1}{2}\int dx\, dt \, dx' \, dt'
V(x-x',t-t') \rho(x,t) \rho(x',t'),
\label{el-el}
\end{equation}
where $V(x,t)$ is the Fourier transform of:
\begin{equation}
V(q,\omega)= \frac{|g(q)|^2}{\zeta}
\frac{1}{\omega^2 - \omega_q^2 + i\eta},
\label{V-phonon}
\end{equation}
being $\zeta$ the ionic mass density and $\omega_q$ the phonon
frequency at momentum $q$. We can distinguish two different situations.
In the so called molecular crystal model (MC), the phonons are optical
and $g(q\to 0)\not = 0$. On the contrary, for acoustic phonons
$\omega_q=c q$  and $g(q)=g \omega_q$.
This is not the most general way of introducing phonons. Another
possibility would be to couple phonons with electrons via a modulation
of the hopping matrix element as in the Su, Schrieffer and Heeger
model\cite{SSH} (SSH) for acoustic phonons.
If we were interested in the low energy
behavior, we could write also in this case
an expression like (\ref{el-el})
with $g(q)=ig\sin(q)$ in (\ref{V-phonon}).

A possible approach to analyze the effects of (\ref{el-el})
close to half filling is that one followed by
Martin and Loss\cite{Martin} which we shall now  discuss in order to
point out the basic limits and approximations.
Following them, for the time being, we limit our
analysis to $V(q,\omega)$ with $q\sim 0$,
and disregard the $q\sim 2k_F$ component. This is in principle an
unrealistic approximation since the $2k_F$-phonons are strongly coupled
to the electrons. However, as we have said in the Introduction, even if
$2k_F$-phonons are taken into account,
the charge modes can still be  described by a Hamiltonian like
(\ref{lutt-liquid}), with renormalized, in particular smaller, parameters
$v_\rho$ and $K_\rho$.
A more relevant point is that the phonon-induced
$q\sim 0$ interaction (\ref{el-el})
introduces in the model new energy scales which is
related to the frequency dependence of
(\ref{el-el}),
and can be taken
as a characteristic phonons frequency.
If we could assume that this energy scale is much smaller than
the energy scale $\omega_L$ below which the Hubbard model
(or better the Hubbard model plus the $2k_F$ lattice distortion)
can be described in terms of a Luttinger liquid,
then it would be justified to proceed along
the lines of Ref.\onlinecite{Martin}
(this energy scale inequality is the crucial
approximation which underlies Martin and Loss's approach).
We therefore first represent the Hubbard model as an effective Luttinger liquid
as in Eq.~(\ref{lutt-liquid}),
with the appropriate parameters $v_\rho$ and $K_\rho$.
This step implies that we are interested in energies $\ll \omega_L$.
Secondly, to this low energy model we add the electron-phonon coupling
of (\ref{el-ph}) limited only to the low momentum component. This
implies that
\[
\rho_\uparrow + \rho_\downarrow \simeq
\sqrt{\frac{2}{\pi}}\nabla \phi_\rho(x)
\]
in (\ref{el-ph}).
If the previous assumption remains valid
close to half-filling, then the effective model can be diagonalized exactly
(see Ref.~\onlinecite{Lucchini}), and describes low momentum
phonons coupled to charge sound waves with vanishing velocity
$v_\rho$, which in turn means that $V$ is always in the strong
coupling regime. A simple consequence of this is that the
charge compressibility should diverge before half-filling:
in fact $\chi^{-1} - v_{\rho}/K_\rho \propto \lim_{q\to 0} V(q,0)$ and
therefore for any weak attraction $V<0$ there is a filling $\nu$
such that $v_{\rho}$ is so small that $\chi^{-1}$ vanishes.
The divergent compressibility identifies the Wentzel-Bardeen
singularity, as discussed by Martin and Loss, or, equivalently,
the instability to phase separation. Notice that this conclusion is
more or less independent of the particularly chosen phonons
(i.e. optical or acoustic) as well as of the
electron-phonon coupling (MC or SSH).

We believe that the above approach to the problem is not correct.
The simple reason is that $\omega_L$ is a function of the density
and vanishes approaching half filling.
In fact the opening of the charge
gap at $\nu=1/2$ is a consequence of the so-called Umklapp scattering.
This process is singular only at half filling. Away from it, the
singularity is cut-off at energies of order
\begin{equation}
\omega_{Um}\simeq v_F(1-2\nu)/a \simeq t(1-2\nu),
\label{omegaU}
\end{equation}
$a$ being the lattice spacing and $t$ the nearest neighbor
hopping. $\omega_{Um}$ depends in general also on $U$ -- Eq.~(\ref{omegaU})
is its weak coupling limit. The consequence is twofold:
\begin{itemize}
\item[(1)] only at energies lower than $\omega_{Um}$
the gapless Luttinger liquid
behavior shows up -- consequently $\omega_L$ should be identified with
$\omega_{Um}$ ;
\item[(2)] since at energies larger than $\omega_{Um}$
the Umklapp processes are important, the closer the system is to
half filling, the larger is the influence of the Umklapp scattering to
the parameters of the effective Luttinger liquid Hamiltonian.
\end{itemize}
Point (1) implies that close to half filling the energy scales introduced by
the electron-phonon interaction become necessarily comparable to the
energy scale $\omega_L$. Therefore it is more correct
to analyze the effect of low momentum phonons on the full Hubbard model
(plus eventually the $2k_F$-phonons)
than on the effective Luttinger liquid model describing
its low energy behavior without $q\sim 0$ phonons.
Notice that analogous considerations apply near $\nu=0$ since $\omega_L$
vanishes approaching zero filling as well,
$\omega_L\simeq v_F/a \simeq t\nu$.
On the other hand, point (2) implies that, in order to
describe correctly the physics of the model close to half filling,
one has to take into account the effects of the Umklapp scattering
processes, both those generated by the on-site repulsion $U$ and
those induced by the electron-phonon coupling.
Umklapp processes will renormalize the phonon-exchanged interaction as well
as they renormalize the direct el-el interactions. This makes it meaningless
to add the bare phonon-exchanged interaction (\ref{el-el}) to the fully
renormalized low-energy Hamiltonian (\ref{lutt-liquid}) while approaching half
filling (or zero filling).

We will now  re-analyze the electron-phonon interaction moving along
the guidelines of the previous discussion.

\section{The Hubbard model near half filling}

We have seen that at the metal-insulator transition, i.e. for $\nu\to 1/2$,
$v_{\rho }\to 0$ while the charge exponent $K_\rho$ jumps
discontinuously from the asymptotic value $1/2$ to 0. A way to
understand these results without invoking the exact Bethe ansatz solution
would be to use the Renormalization
Group (RG)\cite{geology} and analyze the metal-insulator transition within
the context of commensurate-incommensurate transitions.
The RG scaling equations for the Hubbard model
correctly predict that, in the absence of
Umklapp scattering, the charge sector scales to a Luttinger liquid,
while, when present, the Umklapp scattering
scales to strong coupling, which is interpreted
as the opening of a gap. Close to half filling a better approach would be
to use a two cut-off scaling theory. For energy higher than
$\omega_{Um}$ [see (\ref{omegaU})] the correct RG equations are
those in the presence of Umklapp terms, while below
$\omega_{Um}$ these terms are taken to be equal to zero, and the
RG equations correspond to those of a Tomonaga--Luttinger model.
This weak coupling approach has the merit of reproducing the decrease of
$K_\rho$ as $\nu\to 1/2$, but fails in reproducing other important
features. In fact
\begin{itemize}
\item[(1)] it predicts $v_\rho$ to be an increasing function
of the density, while in the reality $v_\rho\to 0$
for $\nu\to 1/2$;
\item[(2)] it predicts $v_\rho$ to be an increasing function
of the interaction, this is not true close to half
filling\cite{Schulz-Hubbard};
\item[(3)] it predicts $K_\rho$ to vanish continuously as $\nu\to 1/2$.
\end{itemize}
Therefore one has to resort to more refined methods than
weak coupling RG which can not reproduce too detailed features when
flowing to strong coupling.

The first aspect which merits to be clarified is why the
point $K_\rho=1/2$ is so important. First of all, this
value corresponds to the limit
$U=\infty$ of the Hubbard model. In this limit the
on-site interaction is the strongest possible, so it clearly constitutes
a lower bound to $K_\rho$, at least approaching half filling
(longer range repulsion would allow lower values).

Secondly, $K_\rho=1/2$ is exactly
the value at the Luther--Emery line, where the charge Hamiltonian in the
presence of Umklapp scattering can be rewritten in terms of
free spinless fermions with a mass term\cite{Luther-Emery}.
At half filling the chemical potential lies in the middle
of the gap, while away
from half filling it crosses the conduction (or valence) band, so that
the particle (hole) occupation of the conduction (valence) band
is proportional to the deviation from half filling.
The standard approach (see e.g. Ref.~\onlinecite{Schulz}) is then to
describe the gapless low-energy effective charge modes by
linearizing the spinless fermion
spectrum around the new Fermi points. Clearly when the
density is very close to half filling, the chemical potential
lies close to the bottom (top) of the conduction (valence)
band, so that the velocity of the gapless modes is very small
$v\sim (1-2\nu)$, correctly reproducing the result of the exact solution.
Moreover, if the starting $K_\rho\not = 1/2$, one can still work
at the Luther--Emery line, the difference $(K_\rho-1/2)$ being accounted
for by an effective interaction between the resulting spinless
fermions\cite{Schulz,Takada}.
However a (weak) interaction among spinless fermions is not effective
at low density; rather its strength vanishes when
$\nu\to 1/2$\cite{Haldane,Schulz}. Hence for
$\nu\to 1/2$ any generic electronic model with a relevant Umklapp tends
asymptotically to the Luther-Emery line\cite{Haldane}.

Going back to the Hubbard model, this in turn implies that close
to $\nu=1/2$ this model behaves as if $U$ were infinite.
This equivalence holds in the sense that it correctly predicts
the asymptotic behavior of the correlation functions,
but it obviously fails in determining $U$-dependent pre-factors.

These results are obtained for unretarded Umklapp scattering. We argue that
they are still valid in the presence of the retarded interaction
induced by the phonons. Specifically we claim that even if the
$2k_F$-phonons are taken into account, the metal-insulator transition at
half-filling can still be described in terms of a model close to
the Luther-Emery line ($K_\rho\to 1/2$).

The immediate consequence is that any additional
interaction (even an attractive interaction induced by low-momentum
acoustic phonons) is relevant only if its bare value
is comparable to the bare bandwidth, and not to the renormalized charge
velocity. This, we believe, more rigorously explains why Martin and
Loss's results for
the onset of the Wentzel-Bardeen singularity are incorrect close to
half filling.

The above conclusions are simple to demonstrate for phonon frequencies
larger than (or comparable to) the electron bandwidth. In this case, in fact,
one can easily map
the original electron-phonon coupled model into a model of
electrons only with unretarded interactions. This amounts to
approximate $V(q,\omega)$ of (\ref{V-phonon}) with its static limit
\[
V(q,\omega)\simeq V(q,0) = -\frac{|g(q)|^2}{\zeta\omega_q^2}.
\]
The strength of this effective attraction has to be compared with
that of the on-site repulsion $U$. If $U + 2V(0,0) - V(2k_F,0)>0$ then
$K_\rho < 1$ and the system still shows dominant density-wave fluctuations.
As a consequence, one recovers at half filling a standard metal-to-insulator
transition. If, on the contrary, that inequality does not hold,
$K_\rho > 1$ and superconducting fluctuations dominate. In this case
there is no commensurate transition at half filling.
As to the spin sector, if $U+V(2k_F,0)>0$ no spin gap opens
and $K_\sigma=1$ (CDW and SDW coexist). In the opposite case the spin modes
acquire a gap.

On the other hand, the reduction to unretarded interactions can be generalized
to include the case when phonon frequencies are smaller (but not too
smaller) than the electron bandwidth and will be discussed  in the following
Sections for weak electron-phonon coupling.

The opposite case of strong electron-phonon coupling and/or
low $2k_F$-phonon frequency  corresponds to the situation
where the $2k_F$-lattice distortion can be described in terms of
amplitude and phase
fluctuations (as if the amplitude of the lattice distortion order parameter
acquires a finite average value). Due to the finite restoring force
acting against the amplitude fluctuations, these latter can be neglected
at zero temperature. The phase fluctuations of the order parameter
are on the contrary massless and they get strongly coupled to the
charge density waves.
Practically the two move together. These collective electronic
and lattice gapless modes are described again by a Luttinger liquid type
of Hamiltonian with velocity $v_\rho$ and exponent $K_{\rho}$ renormalized
to lower values by the electron-phonon coupling\cite{LR&A,Fukuyama}
(see for instance Fukuyama in Ref.~\onlinecite{review}).
The metal-insulator transition
which takes place at half-filling
is driven by an Umklapp term of the same form as in the pure Hubbard
model\cite{notac}.
Therefore even in this limiting situation
the approach to half filling can be still described as in the usual
incommensurate-commensurate transition; the decrease of $K_{\rho}$ induced
by the $2k_F$ processes will indeed increase the relevance of Umklapp
scattering.

\section{Infinite U limit}

To support the picture given in the previous Section and strongly simplify
the analysis of retardation we consider, with some caution,
the $U\to\infty$ limit, since, as we said, it falls in the same universality
class of the model for arbitrary $U$ close to half-filling as far as
charge degrees of freedom are concerned.
The caution is necessary since at $U\to\infty$ only $4k_F$-CDW
fluctuations survive while $2k_F$-ones are suppressed.
This implies that, as the system approaches half-filling with large on-site
repulsion, the $2k_F$-lattice distortion vanishes.

Keeping in mind that taking $U\to\infty$ we disregard interesting properties
which appear only at finite $U$, let us consider how the system approaches
half-filling in this limit. In this case
it is known that the model is equivalent to a model of free spinless fermions
(whose position on the chain correspond to the position of the electrons,
independently of the spin label) plus an Heisenberg model on the squeezed
chain\cite{Ogata&Shiba}.
The Fermi momentum of the spinless fermions, which we will denote
in the following as $k_F^{sf}$, is twice larger than the spinning fermion one,
i.e. $k_F^{sf}=2k_F$. This for instance means that
half filling for the electrons means filling one for
the spinless fermions. Close to this filling, the spinless fermions
almost occupy all the levels of the band: the Fermi energy is
close to the top of the band and the Fermi velocity is very small
(notice the analogy with the Luther-Emery line).
The electron-phonon interaction translates into an interaction between spinless
fermions and phonons. If the phonons are coupled to
the electron density,
the spin sector is unaffected by the phonons
and still describes an Heisenberg model -- no spin gap opens close
to half filling at large $U$.
On the contrary, in the SSH coupling the modulation of
the hopping induces a modulation of the exchange in the effective Heisenberg
model and in turns leads to a spin-Peierls transition with the consequent
opening of a spin gap (which anyway vanishes as $U\to\infty$).
As to the charge sector, we are left with the problem of
spinless fermions (whose electronic energy scale is the bare bandwidth of the
spinning electrons) in the presence of phonon-induced
interaction.  We start by just considering the zero frequency
limit of this interaction, thus ignoring its detailed frequency dependence.
The important scattering processes
are those at transferred momentum $q\sim 0$ and
$q$ of the order of twice the
spinless-fermion Fermi momentum, which implies four times
the spinning-fermion Fermi momentum $q\sim 2k_F^{sf}=4k_F$. The equivalent
$g$-ological model\cite{geology}  has  the effective $g_2$
scattering process given by:
\[
g_2=V(0)-V(4k_F).
\]
Close to filling one for the spinless fermions, i.e. half filling
for the original electrons, $k_F^{sf}=2k_F\to \pi$ therefore
$[V(0)-V(4k_F)]\to [V(0) - V(2\pi)] =0$.
Once again it comes out that any small-$q$ interaction close to half filling
is not very efficient. The same conclusion is obtained in the zero filling
limit.
Notice the crucial importance of formally keeping both the low and high
transferred-momentum electron-phonon coupling.
In the opposite case, in fact, we
would have reached the wrong conclusion that the system is pushed towards
the Wentzel-Bardeen singularity\cite{notad}.

Similarly to what happens for the Heisenberg model with negative
$J_z$ at finite magnetization\cite{Haldane-Heis}, one still expect an
instability to phase-separation but only for strong enough attraction
$|V|\simeq t$, and not for $|V|\simeq v_F/a$.

The physical meaning of this result is indeed simple and predictable.
It states that long wavelength phonons are not relevant close
to a commensurate transition.

One may still wonder about
retardation effects, which we are going to discuss in the next Section.

\section{Retardation Effects}

We now analyze the consequences of retardation for the effective
spinless fermion model which describes the $U\to\infty$ limit of the Hubbard
model discussed in the previous Section\cite{notae}.

We start by considering free spinless fermions coupled to
optical phonons with a momentum independent coupling constant and
arbitrary filling.
Later on we will discuss the case of acoustic phonons.
The frequency dependent effective $g^{eff}_2$ is:
\[
g^{eff}_2(\epsilon_R,\epsilon_L,\omega)
=V(\omega)-V(\epsilon_L-\omega-\epsilon_R),
\]
where
\[
V(\omega)=\frac{g^2}{\zeta} \frac{1}{\omega^2-\omega_0^2},
\]
$\omega_0$ being the phonon frequency [see Eq.~(\ref{V-phonon})].
$\epsilon_R$ and $\epsilon_L$ are the frequencies of
the incoming right and left fermions respectively, and
$\epsilon_R+\omega$ and $\epsilon_L-\omega$ their
outgoing frequencies. The first term on the right hand side is
a pure $g_2$ term (low momentum transferred). The second is
instead a $g_1$ interaction (momentum transferred $\sim 2k_F^{sf}$),
which for spinless fermions can be cast in the form of a
$g_2$ scattering process. If all the fermionic lines are at zero
frequency,  $g^{eff}_2=0$.
However, already at second order a finite and
{\sl repulsive} $g^{eff}_2$ is generated
\begin{equation}
g^{eff}_2= \frac{1}{2\pi v_F}\left(\frac{g^2}{\omega_0^2 \zeta}\right)^2
\ln\left(\frac{\omega_c}{\omega_0}\right).
\label{eff-g}
\end{equation}
Eq.~(\ref{eff-g}) has been obtained for a linearized band with
a bandwidth cut-off $\omega_c$.
At higher orders three kinds of terms are found. The first kind include
all the logarithmic divergent terms which would be there if a
constant $g^{eff}_2$ given by (\ref{eff-g}) were assumed from the beginning.
The second kind consists of all the non divergent diagrams
proportional to powers of $\ln(\omega_c/\omega_0)$. These terms can
be interpreted as the higher order (finite) corrections to (\ref{eff-g}).
The last class of terms include the divergent diagrams generated by the
previous introduced higher order correction to $g^{eff}_2$.
If
\begin{equation}
\frac{1}{2\pi v_F}\left(\frac{g^2}{\omega_0^2 \zeta}\right)
\ln\left(\frac{\omega_c}{\omega_0}\right)\ll 1
\label{condition}
\end{equation}
then one is allowed to neglect all the higher order corrections
to $g^{eff}_2$. It corresponds to a situation where not even the amplitude of
the lattice distortion order parameter develops.
In this case the problem reduces to a simple $g$-ological
model with the repulsive $g^{eff}_2$ Eq.~(\ref{eff-g}). Consequently, the
system of spinless fermions has dominant CDW fluctuations in the
asymptotically low energy regime. For smaller phonon-frequency one has to
use better approximations. A quite useful approach is
the renormalization group, which can be implemented in various
equivalent ways to take into account the retardation
in both spinning and spinless systems
(see e.g.
Refs.~\onlinecite{Bourbonnais,Voit,Solyom-phonon}). All these analyses
show that retardation favors CDW, confirming the simple calculation we
have just described\cite{nota-ritardo}. Applied to
the original $U\to\infty$ Hubbard model it implies that $K_\rho$ gets smaller
than 1/2 when a local interaction with optical phonons is added.

This analysis can be pushed towards filling one (or zero) for the spinless
model (which means half filling for the $U\to\infty$ Hubbard model).
If we calculate in this limit the effective $g_2$, as we have done
to obtain Eq.~(\ref{eff-g}) for intermediate fillings, we find
a $g_2^{eff}$ vanishing as filling zero or one is approached. This in turns
implies that $K_\rho\to 1/2$ (from below) as $\nu\to 1/2$.
The reason of the vanishing $g_2^{eff}$ is simply that the effective
bandwidth (i.e. the interval around the Fermi energy where the
energy dispersion relation can be approximated by a linear dispersion)
vanishes approaching filling zero or one. This implies that the
electron-phonon interaction is practically un-retarded
($\omega_0\gg \omega_c$), and being local in space,
it is zero for Pauli's principle.
In terms of the starting spinning-model, we find once
again that any additional
interaction close to half filling is unimportant unless its strength
is sufficiently large.

What would it change for acoustic phonons?
This case was studied by Chen {\it et al.}\cite{Lucchini}
taking into account only the
small transferred-momentum electron-phonon coupling, which can be
diagonalized exactly. Their results in the spinless case
are analogous to Martin and Loss's\cite{Martin}, which are derived
for spinning fermions, so that
they suffer from the same problems. In fact if we apply the method
of Ref.\onlinecite{Lucchini} to a spinless
fermion model close to filling one (or zero), where the Fermi velocity
is very small (eventually smaller than the phonon velocity), then
we arrive to the conclusions 1) that superconductivity always dominates;
2) that just before filling one (or zero) phase separation occurs which onset
is characterized by a Wentzel-Bardeen singularity.

In order to correctly describe the approach to filling one,
it turns out important to include also the $2k_F^{sf}$-scattering processes,
since in this limit $2k_F^{sf}\to 0$ and one is not allowed to neglect
these processes while keeping the $q\sim 0$ ones.
A simple way, which
neglects the second order correction to these processes previously
discussed and therefore underestimates their effects, is to
approximate the $2k_F^{sf}$-scattering by its
zero frequency limit, approximation valid for weak electron-phonon
coupling and/or $2k_F^{sf}$-phonon frequency  comparable to the bandwidth
(this is surely the case when the density of the spinless fermions
is close to one). This approximation amounts to write
\[
\lim_{q\to 2k_F^{sf}}V(q,\omega)=
\lim_{q\to 2k_F^{sf}}\frac{g^2}{\zeta}\frac{\omega_q^2}{\omega^2 - \omega_q^2}
= - \frac{g^2}{\zeta}.
\]
It is important to realize that these $2k_F^{sf}$-scattering processes
generate both an effective $g_2$ and a so-called $g_4$
interaction\cite{geology,notaf}, so that the effective phonon induced
interaction can be rewritten as
\begin{equation}
g_2(q,\omega) = g_4(q,\omega)= V(q,\omega)-V(2k_F^{sf},0)=
\frac{g^2}{\zeta}\frac{\omega^2}{\omega^2 - \omega_q^2}.
\label{V-ph-spinless}
\end{equation}
The Hamiltonian can be still diagonalized even in the presence of the
$2k_F^{sf}$-scattering processes, which change profoundly the behavior as
compared to that one predicted by using the approximation of
Ref.\onlinecite{Lucchini}. For instance
we do not encounter any Wentzel-Bardeen singularity for vanishing
spinless fermion Fermi velocity. If $v_F$ is the
bare Fermi velocity and $c$ the phonon velocity, we find, for $v_F\ll c$,
a renormalized charge velocity given by
\[
v_F^{eff}=v_F\sqrt{\frac{2\pi \zeta c^2}{2\pi \zeta c^2 +g^2 v_F}}
\]
which is diminished by the coupling with the phonons but yet remains
positive for any $v_F>0$. In fact, analogously to the case of optical phonons,
we again predict dominant CDW-fluctuations. In particular, in terms of the
original parameter, we still find $K_\rho\to 1/2$ from below.
Rigorously speaking, one can not apply this method really close
to filling one when $2k_F^{sf}\to 0$, since it is not anymore correct
to approximate the fermionic band with a linearized dispersion relation.
However we do not expect any singularity in this limit and therefore
we conclude that even in the case of acoustic phonons the metal-insulator
transition is not preceded by any phase separation.

\section{Conclusions}
In the preceding Sections, we have discussed the approach to half filling
of a 1D system of repulsively interacting electrons in the presence of
electron-phonon coupling. We have argued
that the metal-insulator transition which occurs in a pure system at
half-filling can be described, even in the presence of
a strong coupling to the $2k_F$-lattice deformation,
as a standard 1D incommensurate-commensurate transition\cite{Haldane,Schulz}.
By making use of known results for this transition, we have shown that
the coupling constant between the electrons and the low-momentum phonons
(both optical and acoustic) is strongly renormalized downwards close to
half filling.
In particular we find that the renormalized coupling constant vanishes
at least like the charge velocity as the density $\nu$ goes to 1/2.
Our result thus disagrees with that one found by Martin and
Loss\cite{Martin},
according to whom this coupling constant remains finite when $\nu\to 1/2$.
As to the phase diagram of the Hubbard model, they claim that
the insulating phase at half filling is preceded (in density, keeping
fixed all
the other parameters of the Hamiltonian) by a region of phase separation,
which one could imagine as a region of coexistence
of the insulating phase with
the superconductive one which they find for smaller densities.
On the contrary we think that, if an insulating phase does exist at
half-filling, the MIT is still preceded by a region where
charge density wave fluctuations dominate, as in the absence of phonons.
Rather, even if we neglect the coupling with $2k_F$-phonons,
by taking for instance the $U\to\infty$ limit, we find that low-momentum
($q\sim 0$ modulus a reciprocal lattice vector) phonons only provide an
effective {\sl repulsive} interaction close to
half-filling which effect is to diminish the value of the charge exponent
$K_\rho$ from its $U\to\infty$ limit $K_\rho=1/2$ (although, as the
MIT is approached, $K_\rho\to 1/2$ anyway).

\section{Acknowledgments}
It is a special pleasure to acknowledge stimulating discussions
with L. Yu and M. Grilli.


\begin{references}
\bibitem{notaa}In any case, even though the model with vanishing coupling
to $2k_F$-phonons is somewhat unrealistic, one can consider it as well,
at least on a pure theoretical ground.
\bibitem{notac} In particular, apart from the Umkplapp scattering due
to the on-site repulsion, there would be an Umklapp term\cite{LR&A} generated
by coupling the $q=2k_F$ phonons with the $q=2k_F$ CDW, which away from half
filling is not momentum conserving since $4k_F\not=0$ (modulus a reciprocal
lattice vector), but it is momentum conserving at half filling
($4k_F=2\pi$).
\bibitem{notad} Notice that there is no way to escape from the fact
that $V(4k_F)$ is relevant and cannot be disregarded near $4k_F= 0$ or $2\pi$.
\bibitem{notae} The $U\to\infty$ limit considerably simplifies the analysis
firstly because the spin degrees of freedom are explicitly decoupled from
the charge ones, and secondly
because one can disregard the charge scattering at
$2k_F$, and consequently the coupling to the $2k_F$-phonons,
and only consider the $4k_F=2{k_F}^{sf}$ scattering-processes.
\bibitem{nota-ritardo} In particular the RG analysis for spinning fermions
show that the coupling to the $2k_F$ phonons can always be absorbed into an
effective unretarded repulsion in the weak coupling limit. This in turns
justifies the claim we made that the system of interacting fermions close to
half filling is also close to the Luther-Emery line, even if $2k_F$
phonons are taken into account.
\bibitem{notaf} In fact the $2k_F$-scattering
induces a Fermi velocity renormalization which can be interpreted in terms of
so-called $g_4$ processes. This can be simply seen when
we consider optical phonons with frequency much
larger then the electron bandwidth. In this case the phonon induced
interaction
is practically unretarded. Since it is local in space, it is zero for
Pauli's principle ($n(x)^2=n(x)$ for spinless fermions). This implies
that the interaction induced by the $q\sim 0$ phonons must cancel
exactly that one induced by the $q\sim 2k_F$ phonons. Since the former
contains a $g_4$ process, the latter should contain it too.
\end{references}
\end{document}